\documentclass[english,aps,manuscript]{revtex4}
\usepackage[latin1]{inputenc}
\usepackage{babel}

\makeatletter

\providecommand{\LyX}{L\kern-.1667em\lower.25em\hbox{Y}\kern-.125emX\@}

\newcommand{\bee}{\begin{equation}}
\newcommand{\ee}{\end{equation}}
\newcommand{\beea}{\begin{eqnarray}}
\newcommand{\eea}{\end{eqnarray}}

\makeatother
\begin{document}
{\centering \textbf{\Large Brane Worlds in 5D and Warped Compactifications
in IIB}\Large \par}

{\centering \vspace{0.3cm}\par}

{\centering {\large S. P. de Alwis\( ^{\dagger } \) }\large \par}

{\centering \vspace{0.3cm}\par}

{\centering Physics Department, University of Colorado, \\
 Boulder, CO 80309 USA\par}

{\centering \vspace{0.3cm}\par}

{\centering \textbf{Abstract}\par}

{\centering \vspace{0.3cm}\par}

{In computing potentials for moduli in for instance type
IIB string theory in the presence of fluxes and branes a factorisable
ansatz for the ten dimensional metric is usually made. We investigate
the validity of this ansatz by examining the cosmology of a brane
world in a five dimensional bulk and find that it contradicts the
results obtained by using a factorizable ansatz. We explicitly identify
the problem with the latter in the IIB case. These arguments support
our previous work on this question. \par}

\vspace{0.3cm}


\vfill

\( ^{\dagger } \) {\small e-mail: dealwis@pizero.colorado.edu}{\small \par}

\eject

\section{Introduction}

Following the work of Giddings et al \cite{Giddings:2001yu} (GKP)\footnote{This work was based on earlier work by \cite{Polchinski:1995sm}\cite{Becker:1996gj}\cite{Gukov:1999ya}\cite{Gukov:1999gr}\cite{Taylor:1999ii}\cite{Dasgupta:1999ss}\cite{Greene:2000gh}.}
there has been much activity in computing potentials for the moduli
in type IIB theories in the presence of fluxes and D-branes/orientifold
planes \cite{DeWolfe:2002nn}\cite{Kachru:2003aw}\cite{Blumenhagen:2003vr}\cite{Escoda:2003fa}\cite{deAlwis:2003sn}\cite{Berg:2003np}\cite{Buchel:2003js}\cite{Burgess:2003ic}\cite{Brustein:2004xn}\cite{Camara:2003ku}\cite{Grana:2003ek}\cite{Becker:2004gw}\cite{Saltman:2004sn}\cite{Lust:2004fi}.
In all these calculations the following ansatz is made for reducing
the theory to an effective four dimensional one:\begin{equation}
\label{metric}
ds^{2}=g_{MN}dx^{M}dx^{N}=e^{2\omega (y)-6u(x)}\bar{g}_{\mu \nu }(x)+e^{-2\omega (y)+2u(x)}(\tilde{g}_{mn}(y)+z_{i}(x)\phi ^{i}_{mn}(y)+..)dy^{m}dy^{n}.
\end{equation}

In the above \( \tilde{g}_{mn}(y \)) is the metric on the internal
Calabi-Yau manifold, \( u(x) \) is the volume modulus and the \( z_{i}(x) \)
are the other Kahler and complex structure moduli. The question we
wish to address is the following: does the potential obtained by using
this ansatz in the ten-dimensional action, yield the correct four-dimensional
equations on a (3+1 dimensional) brane obtained by projecting the
ten dimensional equations?

In a previous paper \cite{deAlwis:2003sn} we made the following observations.
The no-go theorem, which in effect states that the strong energy condition
is satisfied in the effective four dimensional theory, if it is satisfied
in the ten-dimensional theory (as is the case in string theory) was
shown to be inapplicable here, since the the volume modulus is not
stabilized. It followed from this that to have a consistent derivation
of the potential one needed to keep the time dependence of the volume
modulus. In \cite{deAlwis:2003sn} it was pointed out that doing so,
in the presence of non-trivial warping seemed to require one to keep
all the Kaluza-Klein excitations as well. In effect, while the static
solutions in the presence of fluxes and branes was certainly valid,
the actual form of the 4D potential was not really established. 

In this paper we will highlight the problems associated with the factorized
form of the metric ansatz used in the literature, by considering the
far simpler case of the original \cite{Randall:1999vf} (RS1) construction
of a 3+1 dimensional brane world embedded in five dimensions. We will
see explicitly that the (non-static) five dimensional equations (assuming
homogeneity in three-space) projected on to the brane \cite{Binetruy:1999ut}\cite{Cline:1999ts}\cite{Kanti:1999sz}\cite{Binetruy:1999hy}\cite{Mukohyama:1999qx}\cite{Csaki:1999mp}\cite{Mukohyama:1999wi},
(for recent reviews see \cite{Langlois:2002bb} \cite{Maartens:2003tw}\cite{Brax:2004xh})
are manifestly different from the effective four dimensional equations
obtained from the above factorized metric ansatz. The latter fails
to capture the correct four dimensional physics. Finally we will revisit
the full ten-dimensional theory and discuss precisely where the ansatz
fails, in the light of our five-dimensional investigation.

\section{4D effective action from metric ansatz}

The action for the theory is 

\begin{equation}
\label{fivedaction}
S=\frac{1}{2\kappa ^{2}_{5}}\int d^{5}x\sqrt{g}R^{(5)}-\int d^{5}x\sqrt{g}(\frac{1}{2}(\partial \phi )^{2}+V(\phi ))-\sum _{i=0,\pi }\int d^{4}x\sqrt{g_{i}^{(4)}}L_{i}
\end{equation}

We have included the action for two branes in the above with the energy
density of each brane being split up into a tension part (which in
general will depend on the bulk scalar) and a matter density. The
fifth dimension is taken to be an \( S^{1}/Z_{2} \) orbifold and
we have chosen a gauge such that the branes are located at the fixed
points \( y=0,\pi l_{5},\, (l_{5}^{3}=\kappa _{5}^{2}) \) of the
\( Z_{2} \) action. The metric is taken to be block diagonal,

\[
ds^{2}=g_{\mu \nu }(x,y)dx^{\mu }dx^{\nu }+b^{2}(x,y)dy^{2}\]

If one were to proceed in analogy with what is done in the analysis
of the corresponding type IIB case one would use the metric ansatz\begin{equation}
\label{5metric}
ds^{2}=e^{2A(y)-u(x)}\tilde{g}_{\mu \nu }(x)dx^{\mu }dx^{\nu }+e^{2u(x)}dy^{2}.
\end{equation}
 The salient feature of the above is the assumption of factorisability
of \( x \) and \( y \) dependences (the \( e^{-u} \) factor in
the first term is inserted in order to decouple the modulus field
\( u \) from the four dimensional metric to get the Einstein frame).
One could have of course chosen the \( y \) dependent factor in the
second term to be as in (\ref{metric}) but this would amount to a
trivial redefinition of the coordinate. Inserting this ansatz into
the five dimensional action (\ref{fivedaction}) we get\begin{equation}
\label{4action}
S=\frac{1}{2\kappa ^{2}_{5}}\int dye^{2A(y)}\int d^{4}x\sqrt{\tilde{g}}[\tilde{R}^{(4)}-\frac{3}{2}\widetilde{(\partial _{x}u)^{2}}]-\int d^{4}x\sqrt{\tilde{g}}\int dy\frac{1}{2}e^{2A(y)}\widetilde{(\partial _{x}\phi )^{2}}-\int d^{4}x\sqrt{\tilde{g}}U(u,\phi ).
\end{equation}

The tilde in the above denotes contraction with the tilde metric in
(\ref{5metric}) and the potential \( U \) is given by

\begin{eqnarray}
U(u,\phi )= & -ae^{-3u(x)}+\int dy[\frac{1}{2}e^{4A(y)-3u(x)}(\partial _{y}\phi )^{2}+e^{4A(y)-u(x)}V(\phi )] & \nonumber \\
 & +e^{-2u(x)}\{e^{4A(0)}L_{0}+e^{4A(\pi )}L_{\pi }\}. & \label{4dpot} 
\end{eqnarray}

In the above the constant \( a=\frac{1}{2\kappa _{5}^{2}}\int dy12e^{4A(y)}(\partial _{y}A)^{2}>0 \)
and \( L_{0,\pi } \) is the Lagrangian on each brane. 

Several remarks are in order here. Firstly we have an effective four
dimensional theory very much like the one in the type IIB case of
GKP. There is a potential for the modulus \( u(x) \) that depends
on the bulk scalar field \( \phi  \). The scalar field is the analog
of the four-form field in the GKP case and to make the correspondence
one would also require a static solution to the bulk scalar field
equation to be substituted in to the above. Note also that there is
a critical point for the potential for \( u(x) \) even in the absence
of a bulk scalar field. Finally it is clear that the four dimensional
gravitational equations will be of the usual form, and in particular
the Friedman equation describing four dimensional cosmology will be
the usual one.

This analysis is in sharp conflict with what emerges from the projection
to one or other brane of the five dimensional equations, as we shall
see in the next section.

\section{4D projection of 5D equations}

There is a large literature on brane world cosmology and the relevant
original works were quoted in the introduction. As in those works
we look for spatially homogeneous solutions so the metric is parametrized
as\[
ds^{2}=-n^{2}(t,y)dt^{2}+a^{2}(t,y)(dx_{i})^{2}+b^{2}(t,y)dy^{2}.\]
 Note that we can always use a gauge where \( n(t,y)=b(t,y) \) but
we will not do so here. The gravitational field equations then become\begin{eqnarray}
 & \nonumber \\
 & \nonumber \\
\frac{\dot{a}^{2}}{a^{2}}+\frac{\dot{a}\dot{b}}{ab}-\frac{n^{2}}{b^{2}}\left( \frac{a''}{a}+\frac{a'^{2}}{a^{2}}-\frac{a'b'}{ab}\right)  & =\frac{\kappa _{5}^{2}}{3}(\rho _{0}n^{2}\frac{\delta (y)}{b}+\rho _{\pi }n^{2}\frac{\delta (y-\pi )}{b}+T^{\phi }_{00})\label{00} \\
\frac{a^{2}}{b^{2}}\left[ \frac{a'}{a}\left( \frac{a'}{a}+2\frac{n'}{n}\right) -\frac{b'}{b}\left( \frac{n'}{n}+2\frac{a'}{a}\right) +2\frac{a''}{a}+\frac{n''}{n}\right]  &  & \nonumber \\
+\frac{a^{2}}{n^{2}}\left[ \frac{\dot{a}}{a}\left( -\frac{\dot{a}}{a}+2\frac{\dot{n}}{n}\right) -2\frac{\ddot{a}}{a}+\frac{\dot{b}}{b}\left( -2\frac{\dot{a}}{a}+\frac{\dot{n}}{n}\right) \right]  & =\kappa _{5}^{2}(p_{0}a^{2}_{0}\frac{\delta (y)}{b}+p_{\pi }a_{\pi }^{2}\frac{\delta (y-\pi )}{b}+T_{ii}^{\phi })\label{ii} \\
\frac{n'}{n}\frac{\dot{a}}{a}+\frac{a'}{a}\frac{\dot{b}}{b}-\frac{\dot{a}'}{a} & =\frac{\kappa _{5}}{3}T^{\phi }_{05}\label{05} \\
\frac{a'}{a}\left( \frac{a'}{a}+\frac{n'}{n}\right) -\frac{b^{2}}{n^{2}}\left[ \frac{\dot{a}}{a}\left( \frac{\dot{a}}{a}-\frac{\dot{n}}{n}\right) +\frac{\ddot{a}}{a}\right]  & =\frac{\kappa _{5}^{2}}{3}T^{\phi }_{55}.\label{55} 
\end{eqnarray}
In the above a dot denotes differentiation with respect to time and
a prime with respect to \( y \).

We also have the scalar field equation\[
-\frac{1}{na^{3}b}\partial _{0}(n^{-1}a^{3}b\partial _{0}\phi )+\frac{1}{na^{3}b}\partial _{y}(na^{3}b^{-1}\partial _{y}\phi )=\frac{dV}{d\phi }+\frac{dT_{0}}{d\phi }\frac{\delta (y)}{b}+\frac{dT_{0}}{d\phi }\frac{\delta (y-\pi )}{b}\]

In the above we have split the Lagrangian on the brane into a \( \phi  \)
dependent tension \( T_{0,\pi }(\phi ) \) and a \( \phi  \) independent
matter term \( L_{i}=T_{i}(\phi )+L_{im} \).

Now we may write,\begin{eqnarray}
\ln a(t,y)= & \ln a_{0}(t)+\frac{1}{2}A_{1}(t,y)|y|+A_{2}(t,y) & \label{aansatz} \\
\ln n(t,y)= & \ln n_{0}(t)+\frac{1}{2}N_{1}(t,y)|y|+N_{2}(t,y) & \label{nansatz} \\
\ln b(t,y)= & \ln b_{0}(t)+\frac{1}{2}B_{1}(t,y)|y|+B_{2}(t,y), & \label{bansatz} 
\end{eqnarray}
 where \( A_{i}(t,y)=\sum _{n=0}^{\infty }a_{in}(t)\cos ny=A_{i}(t)+O(y^{2}) \)
with \( A_{2}(t)=0 \), and \( \frac{1}{2}A_{1}(t,\pi )\pi +A_{2}(t,\pi )=\ln \frac{a_{\pi }(t)}{a_{0}(t)} \)
and \( N_{i},\, B_{i} \) satisfy similar relations. We note in passing
that this parametrization imposes no topological constraint on the
function \( A_{1}(t,y) \) - since the integral \( \oint \left( \frac{a'}{a}\right) 'dy \)
is identically zero when the RHS of (\ref{aansatz}) is used to calculate
the integrand. The same remarks are valid for the other functions
\( N_{1,}B_{1} \). 

From eqn (\ref{00}) we get, by integrating over vanishing intervals
around \( y=0 \) and \( y=\pi  \), the boundary conditions\begin{equation}
\label{abdy}
A_{1}(t,0)=\frac{\kappa ^{2}_{5}}{3}\rho _{0}(t)b_{0}(t),\, \, A_{1}(t,\pi )=-\frac{\kappa ^{2}_{5}}{3}\rho _{\pi }(t)b_{\pi }(t).
\end{equation}

Similarly from eqn (\ref{ii}) we get \begin{equation}
\label{nbdy}
N_{1}(t,0)=\frac{\kappa ^{2}_{5}}{3}b_{0}(t)(2\rho _{0}(t)+3p_{0}(t)),\, \, N_{1}(t,\pi )=-\frac{\kappa _{5}^{2}}{3}b_{\pi }(t)(2\rho _{\pi }(t)+3p_{\pi }(t))
\end{equation}
 Also from the scalar field equation we have the matching condition
\begin{equation}
\label{phibdy}
\phi '^{2}_{0,\pi }=\frac{b_{0,\pi }^{2}}{4}\left( \frac{dT_{0,\pi }}{d\phi }\right) ^{2}.
\end{equation}

Note that if we had assumed that \( A_{1}(t,y),\, N_{1}(t,y) \) are
\( y \)-independent (as is done in much of the literature and corresponds
to the RS solution which is valid in the absence of a bulk field)
then we would have been forced to the constraint \begin{equation}
\label{constraint}
\rho _{0}(t)b_{0}(t)=-\rho _{\pi }(t)b_{\pi }(t),\, \, p_{0}(t)b_{0}(t)=-p_{\pi }(t)b_{\pi }(t).
\end{equation}

Of course in the absence of a bulk scalar field one would have the
static RS solution which essentially results in this constraint. Thus
avoidance of this constraint seems to require a bulk field. Note that
our result here is somewhat different from the conclusion of \cite{Csaki:1999mp}
where it is argued that the issue depends on the existence of a stabilization
mechanism for the modulus \( b(t) \). The argument above shows that
the real problem is the assumption of linearity in \( y \) as in
the RS1 solution (where of course it is a consequence of the equations
of motion). 

The boundary conditions eqns (\ref{abdy})(\ref{nbdy}) highlight
the problem with the metric ansatz eqn (\ref{5metric}). They show
that in a dynamic situation the factorization of the metric components
into a \( y \)-dependent and a \( t \) (or \( x \)) dependent factor
is simply not valid, since according to them \( A_{1}(t,y) \) and
\( N_{1}(t,y \)) are necessarily time dependent, whilst the factorization
ansatz would imply that they are purely \( y \)-dependent. It follows
that the equations of motion coming from the effective four dimensional
action (\ref{4action}) will be incorrect in the sense that they would
not be compatible with the 5 dimensional equations of motion except
in the static case.

Let us for instance discuss the analog of the Friedman equation for
this case. This has been discussed at length in the literature beginning
with the work of Binetruy et al (\cite{Binetruy:1999ut}). Nevertheless
for completeness we will re-derive it (especially since most derivations
are done in the absence of a bulk scalar). 

Using eqns (\ref{aansatz})(\ref{nansatz})(\ref{abdy})(\ref{nbdy})
in eqns (\ref{55}) and (\ref{05}) evaluated at \( y=0 \) we get
\begin{eqnarray*}
\frac{\ddot{a}_{0}}{a_{0}}+\frac{\dot{a}_{0}^{2}}{a_{0}^{2}} & = & -\frac{\kappa _{5}^{2}}{3b_{0}^{2}}T_{55}^{\phi }|_{0}-\frac{\kappa ^{2}_{5}}{36}\rho _{0}(\rho _{0}+p_{0})\\
\dot{\rho }_{0}+3(\rho _{0}+p_{0})\frac{\dot{a}_{0}}{a_{0}} & = & 2T_{05}^{\phi }|_{0+}=\frac{2}{b_{0}}\phi '_{0+}\dot{\phi }_{0}
\end{eqnarray*}

Eliminating \( p \) and integrating (after multiplying by an integrating
factor \( a^{4} \)) we get\begin{equation}
\label{F1}
H_{0}^{2}=\frac{\kappa _{5}^{4}}{36}\rho _{0}^{2}-\frac{\kappa _{5}^{4}}{9}a_{0}^{-4}\int \frac{\rho _{0}}{b_{0}}\dot{\phi }_{0}\phi '_{0}a_{0}^{3}da_{0}-\frac{2\kappa ^{2}_{5}}{3}a_{0}^{-4}\int ^{a_{0}}daa_{}^{3}(\frac{\phi '^{2}}{2b_{0}^{2}}+\frac{\dot{\phi }_{0}^{2}}{2}-V_{0})+\frac{\mu }{a_{0}^{4}}
\end{equation}
 where the last term (with \( \mu  \) an integration constant) is
often referred to as 'dark radiation' in the literature. 

This equation looks very different from the usual four-dimensional
one. To see the connection let us first ignore the scalar field and
put \( V=\Lambda  \) a bulk cosmological constant as in RS1 and set
the arbitrary constant \( \mu =0 \). Then we split the energy density
on the brane into a tension piece and a matter piece i.e.\[
\rho _{0}=T_{0}+\rho _{m0},\]

to get\begin{equation}
\label{F2}
H_{0}^{2}=\frac{\kappa _{5}^{4}}{18}T_{0}\rho _{m0}(1+\frac{\rho _{m0}}{2T_{0}})+\frac{\kappa ^{2}_{5}}{6}(\frac{\kappa ^{2}_{5}}{6}T_{0}^{2}+\Lambda ).
\end{equation}

The second term on the RHS is an effective four-dimensional cosmological
constant and to obtain a static solution in the absence of brane matter
one would need to use the RS1 fine-tuning condition \( \frac{\kappa ^{2}_{5}}{6}T_{0}^{2}+\Lambda =0 \).
In any case we see that for \( \rho _{m0}<<T_{0} \), we get the usual
equation provided that we identify the four dimensional gravitational
coupling as \( \kappa _{4}^{2}=\frac{\kappa _{5}^{4}}{6}T_{0} \).
All this is well-known and can be found in several of the papers quoted
in the introduction. However in the absence of a bulk scalar (or some
other equivalent bulk physics) the modulus \( b(t) \) cannot be stabilized
\cite{Goldberger:1999uk}\cite{DeWolfe:1999cp} and will appear as
a zero mass particle coupling with gravitational strength in four
dimensions. Thus if one wants to get a phenomenologically viable brane
world one needs to stabilize the modulus by for instance including
the bulk scalar. 

It is instructive also to consider the static limit of the modified
Friedman equation (\ref{F1}). Putting \( H=\dot{\phi }=0,\, \rho _{m}=0 \)
(so that \( \rho _{0}=T_{0} \))we get after using (\ref{phibdy}),\[
V_{0}=\frac{1}{8}\left( \frac{dT_{0}}{d\phi }|_{0}\right) ^{2}-\frac{\kappa _{5}^{2}}{6}T_{0}^{2}\]

This is the generalization \cite{DeWolfe:1999cp} of the RS fine tuning
condition in the presence of a scalar field. As pointed out by DeWolfe
et al. this equation by it self is not a fine tuning condition, it
just serves to determine the \( \phi _{0} \), but taken in conjunction
with the boundary condition at \( y=\pi  \) the modulus \( b_{0} \)
gets fixed but one fine tuning is required.

The equation (\ref{F1}) thus reduces to equations investigated in
earlier work in the absence of scalar fields as well as to the static
equation in the presence of scalar fields. However there is a peculiar
feature of this equation which appears to defy interpretation in terms
of a four dimensional effective action. This is the fact that the
kinetic term for the scalar field (as well as the \( y \)-derivative
term) appears with a negative sign, although the potential appears
with the right (i.e. positive) sign. This is not a situation like
that of the dilaton (or the volume modulus) which mixes with the graviton
and would appear to come with the wrong sign kinetic term in the original
(string or Jordan) frame. For instance here if we choose the tension
\( T \) independent of \( \phi  \), the effective four dimensional
Newton constant (\( \kappa _{4}^{2}=\frac{\kappa _{5}^{2}}{6}T \))
is constant and the system is already in the Einstein frame. 

It should be noted that in this system the four dimensional Friedman
equation comes from the \( G_{55} \) equation rather than from the
\( G_{00} \) equation which is the five dimensional Friedman eqation.
It is instructive to compare the results of projecting the latter
equation to the brane with the equation obtained from projecting the
\( G_{55} \) equation i.e. (\ref{F1}). We shall do this in the case
that the bulk scalar as well as the radion are stabilized i.e. \( \dot{\phi },\, \dot{b}=0 \)
(with \( b_{0}=1 \)) . In this case also the integral in (\ref{F1})
is trivial and we get (using \( \rho _{0}=T_{0}+\rho _{m} \) and
(\ref{phibdy}))\begin{eqnarray*}
H_{0}^{2}= & \frac{\kappa ^{4}_{5}}{18}T_{0}\rho _{m}(1+\frac{\rho _{m}}{2T_{0}})+\frac{\mu }{a^{4}} & \\
 & \frac{\kappa ^{2}_{5}}{6}(\frac{1}{6}\kappa ^{2}_{5}T_{0}^{2}-\frac{1}{8}\left (\frac{dT_{0}}{d\phi }\right )^{2}+V) & 
\end{eqnarray*}

On the other hand from the \( G_{00} \) equation (\ref{00}) at \( y=0 \)
after using the expansions for the metric functions (\ref{aansatz})(\ref{nansatz})(\ref{bansatz})
as well as (\ref{abdy}) and (\ref{phibdy} we get (again with \( b_{0}=1 \))
\[
H_{0}^{2}=\frac{\kappa ^{4}_{5}}{9}T_{0}\rho _{m}(1+\frac{\rho _{m}}{2T_{0}})+\frac{\kappa ^{2}_{5}}{3}(\frac{1}{6}\kappa ^{2}_{5}T_{0}^{2}+\frac{1}{8}\left (\frac{dT_{0}}{d\phi }\right )^{2}+V)+\frac{\kappa _{5}^{2}}{3}\frac{B_{1}(t,0)\rho }{4}+A''_{2}(t,0)\]

Comparing these two equations shows that although \( B_{1}(t,0) \)
may be set to zero \( A_{2}''(t,0) \) cannot be zero and in fact\[
A''_{2}(t,0)=-\frac{\kappa ^{4}_{5}}{18}T_{0}\rho _{m}(1+\frac{\rho _{m}}{2T_{0}})+\frac{\mu }{a^{4}}-\frac{\kappa ^{2}_{5}}{6}(\frac{1}{6}\kappa ^{2}_{5}T_{0}^{2}+\frac{3}{8}\left (\frac{dT_{0}}{d\phi }\right )^{2}+V).\]
 In other words, the linear approximation \( A_{2}=0 \) (and \( A_{1}(y,t) \)
independent of \( y) \) is invalid except in the static RS case. 

To recapitulate, the cosmology of the brane world is radically different
from that which would arise from dimensional reduction using the metric
ansatz (\ref{5metric}). One would expect similarly that the cosmology
on a brane in type IIB string theory would not be correctly described
by a naive 4D reduction using the ansatz (\ref{metric}). 

Additional differences arise between the projection of the five dimensional
equations and the effective action obtained by using the metric ansatz
(\ref{5metric}) in the expression for the potential for the modulus
\( b \). This is important for it is precisely the analog of this
ansatz that is used in discussions of the derivation of the moduli
potential in the type IIB case. To see this we use eqns (\ref{aansatz})(\ref{nansatz})(\ref{abdy})(\ref{nbdy})
in eqns (\ref{ii}) -2(\ref{55}) (this linear combination is taken
to eliminate \( \ddot{a} \) terms) evaluated at \( y=0 \), to get\begin{equation}
\label{beqn}
\frac{\ddot{b}_{0}}{b_{0}}+(\dot{a},\dot{b},\dot{n}\, terms)=U'(b).
\end{equation}
 \( U(b) \) is an effective potential for the modulus \( b \) and
(after putting \( n_{0}=1 \)) \begin{eqnarray}
U'(b) & = & \kappa ^{2}_{5}(-\frac{1}{a_{0}^{2}}T_{ii}^{\phi }|_{0}+\frac{2}{3b_{0}^{2}}T_{55}^{\phi }|_{0}+\frac{\kappa ^{2}_{5}}{18}\rho _{0}(\rho _{0}+3p_{0})\nonumber \\
 & + & \frac{\kappa _{5}^{2}}{^{6}}(\kappa _{5}^{2}T_{0}^{2}+\frac{BT_{0}}{b_{0}^{2}})+\frac{1}{b_{0}^{2}}(2A_{2}+N_{2}).\label{bforce} 
\end{eqnarray}
 Now specialize to the RS case where there is no scalar field and
\( T_{55}^{\phi }|_{0}=-b_{0}^{2}\Lambda ,\, T_{ii}^{\phi }|_{0}=-\Lambda a_{0}^{2},\, \rho _{0}=T_{0}a_{0}^{2},\, p=-T_{0} \).
The cosmic acceleration is given by,\[
\frac{\ddot{a}_{0}}{a_{0}}+\frac{\dot{a}_{0}^{2}}{a_{0}^{2}}=\frac{\kappa _{5}^{2}}{3}\Lambda -\frac{\kappa _{5}^{4}}{36}T_{0}(T_{0}-3T_{0}).\]
 So we have a static solution if the RS condition \( \Lambda =-\frac{\kappa _{5}^{2}}{6}T_{0}^{2} \)
is imposed. Now let us look at the force on the modulus \( b \) under
the same conditions. Noting that the RS solution implies \( A_{2}=N_{2}=B=0 \)
we get \[
U'(b)=\frac{\kappa _{5}^{2}}{3}(\Lambda +\frac{1}{6}\kappa _{5}^{2}T_{0}^{2})=0,\]
 where the last equality follows from the RS fine-tuning condition.
Thus (as expected) the modulus \( b \) is undetermined in the absence
of a scalar field and implies that the RS theory has a zero mass particle
coupling with gravitational strength. 

However this contradicts what we would find if the metric ansatz (\ref{5metric})
is used to get the effective potential. For in that case, after setting
\( \phi =0 \) and using the RS solution for the warp factor \[
A(y)=-k|y|,\, k=\frac{\kappa _{5}^{2}}{6}T_{0},\, \kappa _{5}^{2}T_{0}^{2}=-6\Lambda ,\, T_{\pi }=-T_{0}<0,\]

we get from (\ref{4dpot})\[
U(u)=\frac{T_{0}}{2}(e^{-4k\pi l_{5}}-1)(e^{-3u(x)}+e^{-u(x)}-2e^{-2u(x)})\]
 so that \[
U'(u)=\frac{T_{0}}{2}(e^{-4k\pi l_{5}}-1)(-3e^{-3u(x)}-e^{-u(x)}+4e^{-2u(x)}),\]

giving an unstable critical point at \( u=0 \)!

Clearly the metric ansatz (\ref{5metric}) gives the wrong physics.
The root of the problem is the assumption of factorisability of the
metric coefficients into \( x \)-dependent and \( y \)-dependent
factors. As can be seen from eqns (\ref{abdy}), (\ref{nbdy}), inserted
into eqns (\ref{aansatz}), (\ref{nansatz}), the metric does not
factorize. Or to put it another way the assumption of factorizability
is inconsistent with the boundary conditions at the branes. This error
in turn gives an incorrect expression for the potential for the radion
modulus which in particular leads to the incorrect result that there
is a critical point even in the absence of a scalar field.

\section{Conclusions: Lessons for IIB}

What lessons can one draw from this exercise for the system that interests
us - type IIB compactified on a Calabi-Yau orientifold with a stack
of D3 branes. The ten dimensional low energy effective action for
this theory (in the Einstein frame with \( 2\kappa _{10}^{2}=1) \)
is \begin{eqnarray*}
S & = & \int d^{10}X\sqrt{-g}\{R-\frac{1}{2\tau _{I}^{2}}\partial _{M}\tau \partial ^{M}\bar{\tau }-\frac{1}{2.3!\tau _{I}}G_{MNP}\bar{G}^{MNP}-\frac{1}{4.5!}\tilde{F}_{MNPQR}\tilde{F}^{MNPQR}\}\\
 &  & +\frac{1}{4i}\int \frac{C_{4}\wedge G_{3}\wedge \bar{G}_{3}}{\tau _{I}}.
\end{eqnarray*}

In the above \( \tau =C_{0}+ie^{-\phi }, \) \( G_{3}=F_{3}-\tau H_{3}, \)
with \( F_{3}=dC_{2} \) and \( H_{3}=dB_{2} \). Also \( \tilde{F}_{5}=F_{5}-\frac{1}{2}C_{2}\wedge H_{3}+\frac{1}{2}B_{2}\wedge F_{3} \)
with the self-duality condition \( \tilde{F}_{5}=*\tilde{F}_{5} \)
being imposed by hand at the level of the equations of motion.

In addition there is the action for the D3 branes and orientifold
3-planes in Einstein frame

\[
S_{loc}=\sum _{i}\left( -\int _{i}d^{4}xT_{3}\sqrt{|g^{(4)}|}+\mu _{3}\int _{i}C_{4}\right) .\]

Here the integrals are taken over the 4D non-compact space at a point
\( i \) in the internal manifold and \( T_{3}=\mu _{3}>0\, (<0) \)
for a D-brane (orientifold plane). The self-duality of the five form
is satisfied by the following ansatz,

\begin{equation}
\label{5an}
\tilde{F}_{5}=\frac{1}{4!}(1+*)\sqrt{\bar{g}_{4}(x)}d\alpha (x,y)\wedge dx^{0}\wedge ...\wedge dx^{3}
\end{equation}

where \( \alpha (x,y) \) is a scalar function. The four dimensional
effective action may now be derived by introducing the metric ansatz
\begin{equation}
\label{metric3}
ds^{2}=e^{2\omega (y)-6u(x)}\tilde{g}_{\mu \nu }(x)dx^{\mu }dx^{\nu }+e^{-2\omega (y)+2u(x)}\tilde{g}_{mn}(x,y)dy^{m}dy^{n}
\end{equation}

with \( \partial _{\mu }\det \tilde{g}_{mn}=0. \) 

The effective potential was derived in \cite{Giddings:2001yu} by
reducing the ten D action using the static version of this ansatz
(i.e with \( u=0 \) and \( \partial _{\mu }g_{mn}=0 \)) and the
expression (the tilde denotes the use of the metric \( \tilde{g} \)
in the inner product) \begin{equation}
\label{potential}
V=\int d^{6}y\sqrt{\tilde{g}^{(6)}}\frac{e^{4\omega -12u}}{24\tau _{I}}\widetilde{|iG_{3}-*_{6}G_{3}|^{2}}
\end{equation}

was obtained. However, except at the minimum of the potential the
static ansatz cannot really be used and immedately leads to the no-go
theorem forbidding positive potentials \cite{Gibbons:1984kp}\cite{deWit:1987xg}\cite{Maldacena:2000mw}
and the resolution, as pointed out in \cite{deAlwis:2003sn}, is to
include time dependence of the volume modulus \( u(x) \). An attempt
was made to include the moduli and a non-trivial warp factor in \cite{DeWolfe:1999cp}
but it was shown in \cite{deAlwis:2003sn} that a consistent derivation
was not possible without including all the Kaluza-Klein (KK) modes.
In fact it was argued that the full ten-dimensional equations with
time dependent moduli (and except at the minimum of the potential
the moduli are necessarily time-dependent) and non-trivial warp factor,
imply that the metric ansatz (\ref{metric3}) is invalid. The argument
in the previous section for the five dimensional theory highlights
this inconsistency. To see this directly in the current context consider
the Bianchi identity for \( \tilde{F}_{5} \). After using the above
metric ansatz it becomes,\begin{equation}
\label{Bi2}
\tilde{\nabla }_{}^{2}\alpha =\frac{i}{12\tau _{I}}e^{8\omega -4u}\widetilde{G_{mnp}*_{6}\bar{G}^{mnp}}+8\widetilde{\partial _{m}\alpha \partial ^{m}\omega }+e^{8\omega -4u}\sum _{i}\mu _{3}\frac{\delta ^{(6)}(y-y^{i})}{\sqrt{\widetilde{g^{(6)}}}}.
\end{equation}

Integrating this over a small ball of radius \( \varepsilon  \) around
the point \( y=y^{i} \) and letting \( \varepsilon \rightarrow 0 \)
we get\begin{equation}
\label{Bi3}
lim_{\varepsilon \rightarrow 0}\oint _{y_{i}}\nabla _{m}\alpha d\sigma ^{m}=e^{8\omega (y^{i})-4u(x)}\mu _{3}
\end{equation}

In particular this equation implies that the function \( \alpha  \)
cannot be independent of space-time since (except at the minimum of
the potential) \( u(x) \) is space-time dependent. Also from the
Einstein equation with the metric ansatz (\ref{metric3}), after using
(\ref{Bi2}) to eliminate the local source term, we get (for more
details see \cite{deAlwis:2003sn}) 

\begin{eqnarray}
\tilde{R}_{\mu \nu }-\frac{1}{2}\tilde{R}^{(4)}\tilde{g}_{\mu \nu } & =-\frac{1}{4}\tilde{g}_{\mu \nu }[\frac{e^{2\omega }}{12\tau _{I}}|iG_{3}-*_{6}G_{3}|^{2}+e^{-4\omega -8u}(\widetilde{\partial _{m}(\alpha -e^{4\omega }))^{2}} & \label{R41} \\
 & +e^{-8u}(\widetilde{\nabla ^{2}}(\alpha -e^{4\omega })+e^{-4\omega }\partial _{m}e^{4\omega }\partial ^{m}(\alpha -e^{4\omega }))]+...,\nonumber 
\end{eqnarray}

the ellipses denoting first order derivative terms. Again integrating
this equation over a ball of radius \( \varepsilon  \) centered at
\( y=y_{i} \) and taking the radius to zero we have,\[
lim_{\varepsilon \rightarrow 0}\oint _{y_{i}}\nabla _{m}\alpha d\sigma ^{m}=lim_{\varepsilon \rightarrow 0}\oint _{y_{i}}\nabla _{m}e^{4\omega (y)}d\sigma ^{m}\]

Comparing with (\ref{Bi3}) we see as expected that the warp factor
cannot be trivial in the presence of a brane and also that consistency
requires \( \partial _{\mu }u(x)=0 \). This in turn is valid only
at the minimum of the potential. Essentially the problem as in the
five-dimensional case is that the factorization ansatz is not valid
in the presence of branes. 

In conclusion then we have shown that the factorized ansatz for getting
an effective action in four dimensions is likely to give an incorrect
result for the moduli potential. At the (global) minimum of the potential
the condition on the fluxes will of course remain unchanged (this
is essentially determined by supersymmetry) so that arguments that
depend only on static solutions to the classical equations (such as
those in \cite{Kachru:2003aw} (KKLT) where the complex structure
moduli and dilaton are integrated out classically) will remain unchanged.
However arguments that depend on the potential (away from the global
minimum as in some of the computations in \cite{Camara:2003ku}\cite{Grana:2003ek}\cite{Saltman:2004sn})
may not be valid. For instance as observed by the authors of the first
two papers the calculation ot the soft scalar masses for none ISD
fluxes from the potential (\ref{potential}) disagree with that obtained
directly from the D-brane action. At such points the volume modulus
\( u \) is time dependent and the arguments of this paper (and \cite{deAlwis:2003sn})
will apply.

\section{Acknowledgments}

I wish to thank Nemanja Kaloper, Raman Sundrum and Ivonne Zavala for
discussions, and Samik DasGupta for collaboration at the early stages
of this work. I'm also grateful to Steve Giddings for letting me know
that he and Maharana have independently come to a similar conclusion
about the validity of the factorized ansatz for the metric. This research
is supported in part by the United States Department of Energy under
grant DE-FG02-91-ER-40672.

\bibliographystyle{apsrev}
\bibliography{myrefs}

\begin{thebibliography}{38}
\expandafter\ifx\csname natexlab\endcsname\relax\def\natexlab#1{#1}\fi
\expandafter\ifx\csname bibnamefont\endcsname\relax
  \def\bibnamefont#1{#1}\fi
\expandafter\ifx\csname bibfnamefont\endcsname\relax
  \def\bibfnamefont#1{#1}\fi
\expandafter\ifx\csname citenamefont\endcsname\relax
  \def\citenamefont#1{#1}\fi
\expandafter\ifx\csname url\endcsname\relax
  \def\url#1{\texttt{#1}}\fi
\expandafter\ifx\csname urlprefix\endcsname\relax\def\urlprefix{URL }\fi
\providecommand{\bibinfo}[2]{#2}
\providecommand{\eprint}[2][]{\url{#2}}

\bibitem[{\citenamefont{Giddings et~al.}(2002)\citenamefont{Giddings, Kachru,
  and Polchinski}}]{Giddings:2001yu}
\bibinfo{author}{\bibfnamefont{S.~B.} \bibnamefont{Giddings}},
  \bibinfo{author}{\bibfnamefont{S.}~\bibnamefont{Kachru}}, \bibnamefont{and}
  \bibinfo{author}{\bibfnamefont{J.}~\bibnamefont{Polchinski}},
  \bibinfo{journal}{Phys. Rev.} \textbf{\bibinfo{volume}{D66}},
  \bibinfo{pages}{106006} (\bibinfo{year}{2002}), \eprint{hep-th/0105097}.

\bibitem[{\citenamefont{DeWolfe and Giddings}(2003)}]{DeWolfe:2002nn}
\bibinfo{author}{\bibfnamefont{O.}~\bibnamefont{DeWolfe}} \bibnamefont{and}
  \bibinfo{author}{\bibfnamefont{S.~B.} \bibnamefont{Giddings}},
  \bibinfo{journal}{Phys. Rev.} \textbf{\bibinfo{volume}{D67}},
  \bibinfo{pages}{066008} (\bibinfo{year}{2003}), \eprint{hep-th/0208123}.

\bibitem[{\citenamefont{Kachru et~al.}(2003)\citenamefont{Kachru, Kallosh,
  Linde, and Trivedi}}]{Kachru:2003aw}
\bibinfo{author}{\bibfnamefont{S.}~\bibnamefont{Kachru}},
  \bibinfo{author}{\bibfnamefont{R.}~\bibnamefont{Kallosh}},
  \bibinfo{author}{\bibfnamefont{A.}~\bibnamefont{Linde}}, \bibnamefont{and}
  \bibinfo{author}{\bibfnamefont{S.~P.} \bibnamefont{Trivedi}}
  (\bibinfo{year}{2003}), \eprint{hep-th/0301240}.

\bibitem[{\citenamefont{Blumenhagen et~al.}(2003)\citenamefont{Blumenhagen,
  Lust, and Taylor}}]{Blumenhagen:2003vr}
\bibinfo{author}{\bibfnamefont{R.}~\bibnamefont{Blumenhagen}},
  \bibinfo{author}{\bibfnamefont{D.}~\bibnamefont{Lust}}, \bibnamefont{and}
  \bibinfo{author}{\bibfnamefont{T.~R.} \bibnamefont{Taylor}},
  \bibinfo{journal}{Nucl. Phys.} \textbf{\bibinfo{volume}{B663}},
  \bibinfo{pages}{319} (\bibinfo{year}{2003}), \eprint{hep-th/0303016}.

\bibitem[{\citenamefont{Escoda et~al.}(2003)\citenamefont{Escoda, Gomez-Reino,
  and Quevedo}}]{Escoda:2003fa}
\bibinfo{author}{\bibfnamefont{C.}~\bibnamefont{Escoda}},
  \bibinfo{author}{\bibfnamefont{M.}~\bibnamefont{Gomez-Reino}},
  \bibnamefont{and} \bibinfo{author}{\bibfnamefont{F.}~\bibnamefont{Quevedo}},
  \bibinfo{journal}{JHEP} \textbf{\bibinfo{volume}{11}}, \bibinfo{pages}{065}
  (\bibinfo{year}{2003}), \eprint{hep-th/0307160}.

\bibitem[{\citenamefont{de~Alwis}(2003)}]{deAlwis:2003sn}
\bibinfo{author}{\bibfnamefont{S.~P.} \bibnamefont{de~Alwis}},
  \bibinfo{journal}{Phys. Rev.} \textbf{\bibinfo{volume}{D68}},
  \bibinfo{pages}{126001} (\bibinfo{year}{2003}), \eprint{hep-th/0307084}.

\bibitem[{\citenamefont{Berg et~al.}(2003)\citenamefont{Berg, Haack, and
  Kors}}]{Berg:2003np}
\bibinfo{author}{\bibfnamefont{M.}~\bibnamefont{Berg}},
  \bibinfo{author}{\bibfnamefont{M.}~\bibnamefont{Haack}}, \bibnamefont{and}
  \bibinfo{author}{\bibfnamefont{B.}~\bibnamefont{Kors}}
  (\bibinfo{year}{2003}), \eprint{hep-th/0312172}.

\bibitem[{\citenamefont{Buchel}(2004)}]{Buchel:2003js}
\bibinfo{author}{\bibfnamefont{A.}~\bibnamefont{Buchel}},
  \bibinfo{journal}{Phys. Rev.} \textbf{\bibinfo{volume}{D69}},
  \bibinfo{pages}{106004} (\bibinfo{year}{2004}), \eprint{hep-th/0312076}.

\bibitem[{\citenamefont{Burgess et~al.}(2003)\citenamefont{Burgess, Kallosh,
  and Quevedo}}]{Burgess:2003ic}
\bibinfo{author}{\bibfnamefont{C.~P.} \bibnamefont{Burgess}},
  \bibinfo{author}{\bibfnamefont{R.}~\bibnamefont{Kallosh}}, \bibnamefont{and}
  \bibinfo{author}{\bibfnamefont{F.}~\bibnamefont{Quevedo}},
  \bibinfo{journal}{JHEP} \textbf{\bibinfo{volume}{10}}, \bibinfo{pages}{056}
  (\bibinfo{year}{2003}), \eprint{hep-th/0309187}.

\bibitem[{\citenamefont{Brustein and de~Alwis}(2004)}]{Brustein:2004xn}
\bibinfo{author}{\bibfnamefont{R.}~\bibnamefont{Brustein}} \bibnamefont{and}
  \bibinfo{author}{\bibfnamefont{S.~P.} \bibnamefont{de~Alwis}}
  (\bibinfo{year}{2004}), \eprint{hep-th/0402088}.

\bibitem[{\citenamefont{Camara et~al.}(2003)\citenamefont{Camara, Ibanez, and
  Uranga}}]{Camara:2003ku}
\bibinfo{author}{\bibfnamefont{P.~G.} \bibnamefont{Camara}},
  \bibinfo{author}{\bibfnamefont{L.~E.} \bibnamefont{Ibanez}},
  \bibnamefont{and} \bibinfo{author}{\bibfnamefont{A.~M.} \bibnamefont{Uranga}}
   (\bibinfo{year}{2003}), \eprint{hep-th/0311241}.

\bibitem[{\citenamefont{Grana et~al.}(2003)\citenamefont{Grana, Grimm, Jockers,
  and Louis}}]{Grana:2003ek}
\bibinfo{author}{\bibfnamefont{M.}~\bibnamefont{Grana}},
  \bibinfo{author}{\bibfnamefont{T.~W.} \bibnamefont{Grimm}},
  \bibinfo{author}{\bibfnamefont{H.}~\bibnamefont{Jockers}}, \bibnamefont{and}
  \bibinfo{author}{\bibfnamefont{J.}~\bibnamefont{Louis}}
  (\bibinfo{year}{2003}), \eprint{hep-th/0312232}.

\bibitem[{\citenamefont{Becker et~al.}(2004)\citenamefont{Becker, Curio, and
  Krause}}]{Becker:2004gw}
\bibinfo{author}{\bibfnamefont{M.}~\bibnamefont{Becker}},
  \bibinfo{author}{\bibfnamefont{G.}~\bibnamefont{Curio}}, \bibnamefont{and}
  \bibinfo{author}{\bibfnamefont{A.}~\bibnamefont{Krause}}
  (\bibinfo{year}{2004}), \eprint{hep-th/0403027}.

\bibitem[{\citenamefont{Saltman and Silverstein}(2004)}]{Saltman:2004sn}
\bibinfo{author}{\bibfnamefont{A.}~\bibnamefont{Saltman}} \bibnamefont{and}
  \bibinfo{author}{\bibfnamefont{E.}~\bibnamefont{Silverstein}}
  (\bibinfo{year}{2004}), \eprint{hep-th/0402135}.

\bibitem[{\citenamefont{Lust et~al.}(2004)\citenamefont{Lust, Reffert, and
  Stieberger}}]{Lust:2004fi}
\bibinfo{author}{\bibfnamefont{D.}~\bibnamefont{Lust}},
  \bibinfo{author}{\bibfnamefont{S.}~\bibnamefont{Reffert}}, \bibnamefont{and}
  \bibinfo{author}{\bibfnamefont{S.}~\bibnamefont{Stieberger}}
  (\bibinfo{year}{2004}), \eprint{hep-th/0406092}.

\bibitem[{\citenamefont{Randall and Sundrum}(1999)}]{Randall:1999vf}
\bibinfo{author}{\bibfnamefont{L.}~\bibnamefont{Randall}} \bibnamefont{and}
  \bibinfo{author}{\bibfnamefont{R.}~\bibnamefont{Sundrum}},
  \bibinfo{journal}{Phys. Rev. Lett.} \textbf{\bibinfo{volume}{83}},
  \bibinfo{pages}{4690} (\bibinfo{year}{1999}), \eprint{hep-th/9906064}.

\bibitem[{\citenamefont{Binetruy
  et~al.}(2000{\natexlab{a}})\citenamefont{Binetruy, Deffayet, and
  Langlois}}]{Binetruy:1999ut}
\bibinfo{author}{\bibfnamefont{P.}~\bibnamefont{Binetruy}},
  \bibinfo{author}{\bibfnamefont{C.}~\bibnamefont{Deffayet}}, \bibnamefont{and}
  \bibinfo{author}{\bibfnamefont{D.}~\bibnamefont{Langlois}},
  \bibinfo{journal}{Nucl. Phys.} \textbf{\bibinfo{volume}{B565}},
  \bibinfo{pages}{269} (\bibinfo{year}{2000}{\natexlab{a}}),
  \eprint{hep-th/9905012}.

\bibitem[{\citenamefont{Cline et~al.}(1999)\citenamefont{Cline, Grojean, and
  Servant}}]{Cline:1999ts}
\bibinfo{author}{\bibfnamefont{J.~M.} \bibnamefont{Cline}},
  \bibinfo{author}{\bibfnamefont{C.}~\bibnamefont{Grojean}}, \bibnamefont{and}
  \bibinfo{author}{\bibfnamefont{G.}~\bibnamefont{Servant}},
  \bibinfo{journal}{Phys. Rev. Lett.} \textbf{\bibinfo{volume}{83}},
  \bibinfo{pages}{4245} (\bibinfo{year}{1999}), \eprint{hep-ph/9906523}.

\bibitem[{\citenamefont{Kanti et~al.}(1999)\citenamefont{Kanti, Kogan, Olive,
  and Pospelov}}]{Kanti:1999sz}
\bibinfo{author}{\bibfnamefont{P.}~\bibnamefont{Kanti}},
  \bibinfo{author}{\bibfnamefont{I.~I.} \bibnamefont{Kogan}},
  \bibinfo{author}{\bibfnamefont{K.~A.} \bibnamefont{Olive}}, \bibnamefont{and}
  \bibinfo{author}{\bibfnamefont{M.}~\bibnamefont{Pospelov}},
  \bibinfo{journal}{Phys. Lett.} \textbf{\bibinfo{volume}{B468}},
  \bibinfo{pages}{31} (\bibinfo{year}{1999}), \eprint{hep-ph/9909481}.

\bibitem[{\citenamefont{Binetruy
  et~al.}(2000{\natexlab{b}})\citenamefont{Binetruy, Deffayet, Ellwanger, and
  Langlois}}]{Binetruy:1999hy}
\bibinfo{author}{\bibfnamefont{P.}~\bibnamefont{Binetruy}},
  \bibinfo{author}{\bibfnamefont{C.}~\bibnamefont{Deffayet}},
  \bibinfo{author}{\bibfnamefont{U.}~\bibnamefont{Ellwanger}},
  \bibnamefont{and} \bibinfo{author}{\bibfnamefont{D.}~\bibnamefont{Langlois}},
  \bibinfo{journal}{Phys. Lett.} \textbf{\bibinfo{volume}{B477}},
  \bibinfo{pages}{285} (\bibinfo{year}{2000}{\natexlab{b}}),
  \eprint{hep-th/9910219}.

\bibitem[{\citenamefont{Mukohyama}(2000)}]{Mukohyama:1999qx}
\bibinfo{author}{\bibfnamefont{S.}~\bibnamefont{Mukohyama}},
  \bibinfo{journal}{Phys. Lett.} \textbf{\bibinfo{volume}{B473}},
  \bibinfo{pages}{241} (\bibinfo{year}{2000}), \eprint{hep-th/9911165}.

\bibitem[{\citenamefont{Csaki et~al.}(2000)\citenamefont{Csaki, Graesser,
  Randall, and Terning}}]{Csaki:1999mp}
\bibinfo{author}{\bibfnamefont{C.}~\bibnamefont{Csaki}},
  \bibinfo{author}{\bibfnamefont{M.}~\bibnamefont{Graesser}},
  \bibinfo{author}{\bibfnamefont{L.}~\bibnamefont{Randall}}, \bibnamefont{and}
  \bibinfo{author}{\bibfnamefont{J.}~\bibnamefont{Terning}},
  \bibinfo{journal}{Phys. Rev.} \textbf{\bibinfo{volume}{D62}},
  \bibinfo{pages}{045015} (\bibinfo{year}{2000}), \eprint{hep-ph/9911406}.

\bibitem[{\citenamefont{Mukohyama et~al.}(2000)\citenamefont{Mukohyama,
  Shiromizu, and Maeda}}]{Mukohyama:1999wi}
\bibinfo{author}{\bibfnamefont{S.}~\bibnamefont{Mukohyama}},
  \bibinfo{author}{\bibfnamefont{T.}~\bibnamefont{Shiromizu}},
  \bibnamefont{and} \bibinfo{author}{\bibfnamefont{K.-i.} \bibnamefont{Maeda}},
  \bibinfo{journal}{Phys. Rev.} \textbf{\bibinfo{volume}{D62}},
  \bibinfo{pages}{024028} (\bibinfo{year}{2000}), \eprint{hep-th/9912287}.

\bibitem[{\citenamefont{Langlois}(2003)}]{Langlois:2002bb}
\bibinfo{author}{\bibfnamefont{D.}~\bibnamefont{Langlois}},
  \bibinfo{journal}{Prog. Theor. Phys. Suppl.} \textbf{\bibinfo{volume}{148}},
  \bibinfo{pages}{181} (\bibinfo{year}{2003}), \eprint{hep-th/0209261}.

\bibitem[{\citenamefont{Maartens}(2003)}]{Maartens:2003tw}
\bibinfo{author}{\bibfnamefont{R.}~\bibnamefont{Maartens}}
  (\bibinfo{year}{2003}), \eprint{gr-qc/0312059}.

\bibitem[{\citenamefont{Brax et~al.}(2004)\citenamefont{Brax, van~de Bruck, and
  Davis}}]{Brax:2004xh}
\bibinfo{author}{\bibfnamefont{P.}~\bibnamefont{Brax}},
  \bibinfo{author}{\bibfnamefont{C.}~\bibnamefont{van~de Bruck}},
  \bibnamefont{and} \bibinfo{author}{\bibfnamefont{A.-C.} \bibnamefont{Davis}}
  (\bibinfo{year}{2004}), \eprint{hep-th/0404011}.

\bibitem[{\citenamefont{Goldberger and Wise}(1999)}]{Goldberger:1999uk}
\bibinfo{author}{\bibfnamefont{W.~D.} \bibnamefont{Goldberger}}
  \bibnamefont{and} \bibinfo{author}{\bibfnamefont{M.~B.} \bibnamefont{Wise}},
  \bibinfo{journal}{Phys. Rev. Lett.} \textbf{\bibinfo{volume}{83}},
  \bibinfo{pages}{4922} (\bibinfo{year}{1999}), \eprint{hep-ph/9907447}.

\bibitem[{\citenamefont{DeWolfe et~al.}(2000)\citenamefont{DeWolfe, Freedman,
  Gubser, and Karch}}]{DeWolfe:1999cp}
\bibinfo{author}{\bibfnamefont{O.}~\bibnamefont{DeWolfe}},
  \bibinfo{author}{\bibfnamefont{D.~Z.} \bibnamefont{Freedman}},
  \bibinfo{author}{\bibfnamefont{S.~S.} \bibnamefont{Gubser}},
  \bibnamefont{and} \bibinfo{author}{\bibfnamefont{A.}~\bibnamefont{Karch}},
  \bibinfo{journal}{Phys. Rev.} \textbf{\bibinfo{volume}{D62}},
  \bibinfo{pages}{046008} (\bibinfo{year}{2000}), \eprint{hep-th/9909134}.

\bibitem[{\citenamefont{Gibbons}(1984)}]{Gibbons:1984kp}
\bibinfo{author}{\bibfnamefont{G.~W.} \bibnamefont{Gibbons}}
  (\bibinfo{year}{1984}), \bibinfo{note}{three lectures given at GIFT Seminar
  on Theoretical Physics, San Feliu de Guixols, Spain, Jun 4-11, 1984}.

\bibitem[{\citenamefont{de~Wit et~al.}(1987)\citenamefont{de~Wit, Smit, and
  Hari~Dass}}]{deWit:1987xg}
\bibinfo{author}{\bibfnamefont{B.}~\bibnamefont{de~Wit}},
  \bibinfo{author}{\bibfnamefont{D.~J.} \bibnamefont{Smit}}, \bibnamefont{and}
  \bibinfo{author}{\bibfnamefont{N.~D.} \bibnamefont{Hari~Dass}},
  \bibinfo{journal}{Nucl. Phys.} \textbf{\bibinfo{volume}{B283}},
  \bibinfo{pages}{165} (\bibinfo{year}{1987}).

\bibitem[{\citenamefont{Maldacena and Nunez}(2001)}]{Maldacena:2000mw}
\bibinfo{author}{\bibfnamefont{J.~M.} \bibnamefont{Maldacena}}
  \bibnamefont{and} \bibinfo{author}{\bibfnamefont{C.}~\bibnamefont{Nunez}},
  \bibinfo{journal}{Int. J. Mod. Phys.} \textbf{\bibinfo{volume}{A16}},
  \bibinfo{pages}{822} (\bibinfo{year}{2001}),
  \eprint[http://arXiv.org/abs]{hep-th/0007018}.

\bibitem[{\citenamefont{Polchinski and Strominger}(1996)}]{Polchinski:1995sm}
\bibinfo{author}{\bibfnamefont{J.}~\bibnamefont{Polchinski}} \bibnamefont{and}
  \bibinfo{author}{\bibfnamefont{A.}~\bibnamefont{Strominger}},
  \bibinfo{journal}{Phys. Lett.} \textbf{\bibinfo{volume}{B388}},
  \bibinfo{pages}{736} (\bibinfo{year}{1996}), \eprint{hep-th/9510227}.

\bibitem[{\citenamefont{Becker and Becker}(1996)}]{Becker:1996gj}
\bibinfo{author}{\bibfnamefont{K.}~\bibnamefont{Becker}} \bibnamefont{and}
  \bibinfo{author}{\bibfnamefont{M.}~\bibnamefont{Becker}},
  \bibinfo{journal}{Nucl. Phys.} \textbf{\bibinfo{volume}{B477}},
  \bibinfo{pages}{155} (\bibinfo{year}{1996}),
  \eprint[http://arXiv.org/abs]{hep-th/9605053}.

\bibitem[{\citenamefont{Gukov et~al.}(2000)\citenamefont{Gukov, Vafa, and
  Witten}}]{Gukov:1999ya}
\bibinfo{author}{\bibfnamefont{S.}~\bibnamefont{Gukov}},
  \bibinfo{author}{\bibfnamefont{C.}~\bibnamefont{Vafa}}, \bibnamefont{and}
  \bibinfo{author}{\bibfnamefont{E.}~\bibnamefont{Witten}},
  \bibinfo{journal}{Nucl. Phys.} \textbf{\bibinfo{volume}{B584}},
  \bibinfo{pages}{69} (\bibinfo{year}{2000}), \eprint{hep-th/9906070}.

\bibitem[{\citenamefont{Gukov}(2000)}]{Gukov:1999gr}
\bibinfo{author}{\bibfnamefont{S.}~\bibnamefont{Gukov}},
  \bibinfo{journal}{Nucl. Phys.} \textbf{\bibinfo{volume}{B574}},
  \bibinfo{pages}{169} (\bibinfo{year}{2000}), \eprint{hep-th/9911011}.

\bibitem[{\citenamefont{Taylor and Vafa}(2000)}]{Taylor:1999ii}
\bibinfo{author}{\bibfnamefont{T.~R.} \bibnamefont{Taylor}} \bibnamefont{and}
  \bibinfo{author}{\bibfnamefont{C.}~\bibnamefont{Vafa}},
  \bibinfo{journal}{Phys. Lett.} \textbf{\bibinfo{volume}{B474}},
  \bibinfo{pages}{130} (\bibinfo{year}{2000}),
  \eprint[http://arXiv.org/abs]{hep-th/9912152}.

\bibitem[{\citenamefont{Dasgupta et~al.}(1999)\citenamefont{Dasgupta, Rajesh,
  and Sethi}}]{Dasgupta:1999ss}
\bibinfo{author}{\bibfnamefont{K.}~\bibnamefont{Dasgupta}},
  \bibinfo{author}{\bibfnamefont{G.}~\bibnamefont{Rajesh}}, \bibnamefont{and}
  \bibinfo{author}{\bibfnamefont{S.}~\bibnamefont{Sethi}},
  \bibinfo{journal}{JHEP} \textbf{\bibinfo{volume}{08}}, \bibinfo{pages}{023}
  (\bibinfo{year}{1999}), \eprint{hep-th/9908088}.

\bibitem[{\citenamefont{Greene et~al.}(2000)\citenamefont{Greene, Schalm, and
  Shiu}}]{Greene:2000gh}
\bibinfo{author}{\bibfnamefont{B.~R.} \bibnamefont{Greene}},
  \bibinfo{author}{\bibfnamefont{K.}~\bibnamefont{Schalm}}, \bibnamefont{and}
  \bibinfo{author}{\bibfnamefont{G.}~\bibnamefont{Shiu}},
  \bibinfo{journal}{Nucl. Phys.} \textbf{\bibinfo{volume}{B584}},
  \bibinfo{pages}{480} (\bibinfo{year}{2000}), \eprint{hep-th/0004103}.

\end{thebibliography}

\end{document}